Observation of plastoferrite character and semiconductor to metal transition in soft ferromagnetic $Li_{0.5}Mn_{0.5}Fe_2O_4$ ferrite

R. N. Bhowmik[*]

Department of Physics, Pondicherry University, R.Venkataraman Nagar, Kalapet,

Puducherry - 605 014, India

[*]Corresponding author: Tel.: +91-9944064547; Fax: +91-413-2655734
E-mail: rnbhowmik.phy@pondiuni.edu.in

We prepared $Li_{0.5}Mn_{0.5}Fe_2O_4$ ferrite through chemical reaction in highly acidic solution and subsequent sintering of chemical routed powder at temperatures ≥ 800 $^0$C. Surface morphology showed plastoferrite character for sintering temperature > 1000 $^0$C. Mechanical softening of metal-oxygen bonds at higher measurement temperatures stimulated delocalization of charge carriers, which were strongly localized in A and B sites of the spinel structure at lower temperatures. The charge delocalization process has activated semiconductor to metallic transition in ac conductivity curves, obeyed by Jonscher power law and Drude equation, respectively. Metallic state is also confirmed by the frequency dependence of dielectric constant curves.

Spinel ferrites are attractive due to enriched magnetic and electrical properties, which can be controlled by synthesis route and existence of different metal ions at tetrahedral (A) and octahedral (B) sites of spinel structure [1, 2]. $LiFe_2O_4$ ferrite is a soft ferromagnet, generally used as battery material [3]. $MnFe_2O_4$ ferrite is another ferromagnet applied in power circuits [4]. The mixed compound of $LiFe_2O_4$ and $MnFe_2O_4$ has promised potential applications in microelectronics and information storage devices due to soft ferromagnetic properties with excellent square shaped loop, low electrical conductivity and low energy loss [4-6]. Li-Mn ferrite ($Li_{0.5}Fe_{2.5-x}Mn_xO_4$) has marked many interesting ferromagnetic and electrical properties, which can be controlled by varying Fe and Mn ratio [3, 5, 7]. Here, we report on $Li_{0.5}Mn_{0.5}Fe_2O_4$, a specific compound of $Li_{0.5}Fe_{2.5-x}Mn_xO_4$ series with Fe and Mn ratio 1:1.

The material was prepared by chemical reaction in highly acidic medium. The stoichiometric amount of MnO, $Li_2CO_3$ and $FeCl_3$ were dissolved in concentrated HCl and a solution of pH value ~0.65 was prepared. The mixed solution was heated at 80-90 $^o$C for 18 hrs in stirring condition, and finally, heated at 200 $^o$C for 3 hrs and black color powder was obtained. The powder in pellet form was sintered at different temperatures in air atmosphere to obtain single phased compound. Structural phase of the sintered samples after cooling to room temperature was studied using X-ray diffraction (XRD) pattern in the 2θ range 20-80$^o$ with step size 0.02$^o$ using CuKα radiation. Profile fit of the XRD pattern using Rietveld program, as in Fig. 1, confirmed single phased cubic spinel structure at sintering temperature ≥ 800 $^o$C for 6 hours. The samples sintered at 800 $^o$C, 900 $^o$C, 1000 $^o$C, and 1100 $^o$C were denoted as S8, S9, S10, and S11, respectively. Lattice parameter (8.322-8.36 Å) in the samples increased with sintering temperature as an effect of ordering of cations in A and B sites of cubic spinel structure. Room temperature magnetic measurement confirmed soft ferromagnetic nature of the samples with coercivity 16-68 Oe and spontaneous magnetization 34-58 emu/g. Lattice parameter and ferromagnetic parameters in the present Li-Mn ferrite are consistent to reports [5, 7]. Apart from coexistence of multivalent cations ($Li^+$, $Mn^{2+}$, $Mn^{3+}$, $Fe^{2+}$, $Fe^{3+}$), instability in the ordering of cations in A and B sites greatly modify conduction mechanism in ferrites. Especially, activation of strong lattice vibration and electron-phonon interactions play significant role on controlling the high temperature conduction mechanism. Except our preliminary work [8], there is no report of frequency activated metallic behavior at high temperature conductivity curves of ferrite. We demonstrate this issue and its possible mechanism using ac conductivity study of $Li_{0.5}Mn_{0.5}Fe_2O_4$ ferrite.

Fig.2 shows the surface morphology (SEM) of $Li_{0.5}Mn_{0.5}Fe_2O_4$ samples at different stages of sintering. SEM images showed spherical shaped (lower sintering temperature: Fig. 2(a-c)) and polygonal shaped (higher sintering temperature: Fig. 2(d-e)) particles, whose size increases with sintering temperature. Length and breadth of the particles varied in the range 0.2-0.5 µm and 0.1-0.3µm, respectively. However, grain (crystallite) size of the samples obtained by analyzing XRD peaks using Debye Scherrer formula is in the range 30-50 nm. This means particle size observed from SEM pictures is poly-grained (crystalline) structure. Small sized particles are melted across the interfaces to form micron sized particles and the particles form network structure (Fig. 2(b-d)). It appears that mechanical softness of the particles increases for the samples sintered at higher temperature and large number of small particles are accommodated inside a big sized particle (Fig. 2d). The sample sintered at 1100 $^o$C showed ring shaped or wavy patterned surface. Such unique feature is expected in a special class of material, known as plastoferrite, which is a composite of ferrite particles and organic materials (synthetic rubber, plastic, etc). In fact, plastoferrite feature (thermal activated surface strains) has been seen in many ceramics due to inhomogeneous grain growth process [9, 10]. Due to heterogeneities in system, some parts may be strong enough to resist local stress during cooling process and other parts become sufficiently soft so that displaced atoms can rearrange plastically during cooling process. This develops elasto-plastic constituents in the material. The elasto-plastic effect is prominent at the interfaces (Fig. 2e) where broken bonds, local atomic displacements and effects of intra-grains porosity are easily available. Recent interests for the plastoferrites are growing for understanding the effects of thermal induced visco-elasticity or plasticity on electro-magnetic properties or stress induced deformation in complex network system [11]. Energy dispersive analysis of X-ray (EDX) spectrum (e.g., Fig. 2(f)) indicated elemental composition (Mn:Fe= 0.60:2) of the samples close to expected value 0.5:2 in $Li_{0.5}Mn_{0.5}Fe_2O_4$. Li is not detected in the EDX spectrum due to low atomic number (Z), but its presence was indicated in FTIR spectrum. EDX data indicated slight oxygen deficiency in the samples with oxygen content 4-$\delta$ ($\delta \sim$ 0.2). We correlate a small weight loss ($\leq$ 2.5 %) in TGA curves above 800 $^o$C to the oxygen loss and increasing porosity that resulted in surface deformation in the samples sintered above 1000 $^o$C [12].

Frequency (1 Hz- 5 MHz) dependence of ac conductivity has been studied for two samples with smaller grain size ($\sim$ 30 nm and 35 nm for S8 and S9, respectively). The measurement was performed at field amplitude 1 Volt in the temperature range 300-923 K using broadband dielectric spectrometer (Novo Control Technology, Germany). Disc-shaped samples with ø$\sim$ 12 mm and thickness $\sim$ 2-3 mm were sandwiched between two platinum plates that were connected to the spectrometer. Fig. 3(a-b) shows frequency ($\nu$) dependent ac conductivity ($\sigma_n(\nu)$) (normalized by conductivity at 1Hz) at different temperatures. The immediate observation is that $\sigma_n(\nu)$ is nearly frequency independent ($\sigma_{dc}$ limit) up to certain frequency (say, $\nu_C$). The ac conductivity is rapidly activated at higher frequencies ($\nu > \nu_C$). The remarkable feature is the observation of metallic character in ac conductivity curves above semiconductor to metallic transition temperature ($T_{SM}$) $\geq$ 740 K. In semiconductor regime (T <$T_{SM}$), ac conductivity increases with the increase of frequency. In metallic regime (T >$T_{SM}$), ac conductivity decreases with the increase of frequency. The semiconductor and metallic regimes are magnified in Fig. 3(c-d). This semiconductor to metallic transition is different from the semiconductor to metallic transition reported in other ferrites, e.g., NiFe$_2$O$_4$ at $\sim$ 335 K [13], and CoFe$_2$O$_4$ at $\sim$ 330-450 K [14-15]. First, metallic conductivity in these ferrites was noted in the temperature dependence of grain boundary conductivity data, which was derived from impedance spectra. Second, metallic feature was not directly observed in ac conductivity curves, which may be related to not sufficiently high measurement temperatures or not intrinsic properties of the samples. In the present Li-Mn ferrite, metallic response is directly observed in ac conductivity curves within measurement temperatures. We noted such semiconductor to metallic transition and plastoferrite character not only in chemical routed samples, but also in solid state routed samples with micron-sized grains [8]. Hence, metallic conductivity in present ferrite is not directly related to either grain size effect or structural disorder (porosity, lattice defects), of course grain size and structural disorder influence up to certain extent. Now, we demonstrate the mechanism in semiconductor and metallic regimes of the samples.

The frequency activated increase of conductivity [$\sigma_{ac}(\nu, T < T_{SM})$] in semiconductor regime obeys Jonscher's power law [16]: $\sigma_{ac}(\nu,T) = \alpha(T)\nu^n$ with temperature dependent parameter $\alpha(T)$ and n, a dimension less parameter. The exponent n has been calculated from the fit of conductivity curves (shown in magnified figures of Fig. 3(c-d)) for $T < T_{SM}$. As shown in left hand side of Fig. 4(a), the obtained values of n are less than 1 and decreases with increasing temperature up to 625 K. In the absence of free charge carriers, frequency activated conductivity in semiconductor regime of $Li_{0.5}Mn_{0.5}Fe_2O_4$ ferrite is explained by hopping of localized charge carriers between B sites ions (electrons between $Fe^{2+}$ and $Fe^{3+}$ and holes between $Mn^{3+}$ and $Mn^{2+}$) [5]. Following earlier report [15], n value smaller than 1 suggests long range hopping of charge carriers and decrease of n with increasing temperature indicates correlated barrier hopping of charge carriers. Above 625 K, n fluctuates as the temperature approaches towards $T_{SM}$ (metallic regime). In this mixed regime of conductivity, n values are affected due to increasing delocalization of charge carriers. In metallic regime of nanostructured materials having localized charge carriers [17], the ac conductivity ($\sigma^*(\omega = 2\pi\nu)$) has been described by Drude-Smith model [18]: $\sigma^*(\omega) = \frac{\sigma_0}{1-i\omega\tau}[1 + \sum_j \frac{c_j}{(1-i\omega\tau)^j}]$. In our samples, ac conductivity ($\sigma'(\omega)$) curves in metallic regime ($T > T_{SM}$) are well fitted according to Drude equation [19]: $\sigma'(\omega) = \sigma_0/(1+\omega^2\tau_s^2)$. This is a special case of Drude-Smith model for isotropic scattering in free electrons system. Here, $\tau_s$ is the relaxation time (collision time between two scattering events, usually electron-phonon scattering) and $\sigma_0$ is the dc limit of conductivity ($\sigma_{dc}$) as we used for normalization of ac conductivity curves. The right hand side of Fig. 4(b) shows that relaxation time ($\tau_s$) of the samples in metallic regime that obtained through fitting of $\sigma^/(\nu)$ data (see Fig. 3(d)). The relaxation time in metallic regime increases with the increase of measurement temperature, which is understood as an effect of increasing electron-phonon interactions. Remarkable feature is that magnitude of $\tau_s$ ($10^{-8}$ -$10^{-9}$ s) in metallic regime is few order less than the relaxation time ($10^{-13}$ -$10^{-15}$ s) for scattering of free electrons usually occurs in optical frequency range. Hence, present ferrite represents a unique system where optical conduction resonance has shifted to lower frequencies [17]. One reason is that charge carriers are not completely free while moving inside the crystal. The ions are either free to move within limited range or executing vibration about their mean positions so that electrons clouds from neighbouring atoms can directly overlap and provide continuous path for charge motion. Goodenough [20] demonstrated that semiconductor feature is expected in oxides when superexchange (cation-anion-cation) interactions dominate over direct (cation-cation) exchange interactions; otherwise metallic feature dominates in the case of strong cation-cation interactions. In this ferrite, localization of charge carries at A and B sites at low temperatures exhibits semiconductor feature. At higher temperatures, probability of transferring electrons between ions at adjacent sites increases due to softening of elasticity or increasing plasticity in metal-oxygen bonds [11]. Fig. 4(b-c) shows a systematic change in conductivity regime with increasing temperature. As indicated in Fig. 4(c), conductivity increases with the increase of frequencies in semiconducting regime and decreases in metallic regime and nearly frequency independent in mixed regime of conductivity. The frequency dependence of dielectric constant curves in metallic regime supports the correlation between dielectric constant ($\varepsilon^*(\omega) = \varepsilon^/(\omega) + i\varepsilon^{//}(\omega)$) and ac conductivity ($\sigma^*(\omega) = \sigma^/(\omega) + i\sigma^{//}(\omega)$), as established in electromagnetic theory using the relations $\varepsilon^*(\omega) -1 = i\sigma^*(\omega)/\varepsilon_0\omega$. This separates out the real and imaginary parts as $\varepsilon'(\omega) - 1 = -\sigma_0\tau_s/\varepsilon_0(1+\omega^2\tau_s^2)$ and $\varepsilon''(\omega) = \sigma_0/\varepsilon_0\omega(1+\omega^2\tau_s^2)$. Fig. 5(a-b) shows that $\varepsilon'$ is positive throughout the frequency range at lower temperatures (semiconductor regime). As metallic feature dominates at higher temperatures, $\varepsilon'$ becomes increasingly negative on lowering the frequency. On the other hand, $\varepsilon''(\nu)$ in Fig. 5 (c-d) remained positive throughout the frequencies for all measurement temperatures. The decrease of $\varepsilon'(\nu)$ with negative magnitude and increase of $\varepsilon''(\nu)$ with positive magnitude as $\nu \rightarrow 0$ confirms the metallic character in a material [17].

We conclude that present system is not a typical metal, but metallic feature appeared in the semiconductor ferrite due to increasing charge delocalization effect in bound ions at A and B sites of the cubic spinel structure. At higher temperatures, the development of semi-elastic properties makes the metal oxygen (Li-O, Fe-O, Mn-O) bonds mechanically soft and flexible for overlapping electronic orbitals from adjacent ions and delocalization of electrons increases. This process is accelerated by rapid vibration of metals (Li, Mn, Fe) ions at higher frequencies of ac field. This helps to break some of the Li-O bonds and produces $Li^+$ ions that move into interstitial sites. In the presence of oxygen vacancy ($Fe^{3+}$-□-$Fe^{2+}$, $Mn^{3+}$-□-$Mn^{2+}$, $Li^+$-□-$Fe^{3+}$), charge carriers in localized state (semiconductor) of superexchange paths ($Fe^{3+}$-O-$Fe^{3+}$, $Mn^{3+}$-O-$Mn^{2+}$, $Li^+$-O-$Fe^{3+}$) transform to delocalized state (metallic). The plastoferrite character may not be directly related to metallic conductivity, but it indicates the materials where one expect greater probability of delocalized charge carriers at higher temperatures. Such ferrites can be used for correlating thermal strain induced structural deformation and relaxation in lattice dynamics. The important technological aspect is that same material can be used as a semiconductor that allows electromagnetic wave to pass through the material at low temperature, but behave as a metallic reflector at high temperature.

Author thanks CIF, Pondicherry University, for providing experimental facilities. Author thanks to Miss. G. Vijayasri for assisting in experimental work.

[1] Y. Zhang, Z. Yang, D. Yin, Y. Liu, C. L. Fei, R. Xiong, J. Shi, and G. L.Yan, J. Magn. Magn. Mater. **322**, 3470 (2010).

[2] M. Srivastava, A.K. Ojha, S. Chaubey, P.K. Sharma, and A.C. Pandey, Mater. Sci. Eng. B **175**, 14 (2010).

[3] M. Gracia, J. F. Marco, J. R. Gancedo, J.L. Gautier, E.I. Ríos, N. Menéndez, and J. Tornero, J. Mater. Chem. **13**, 844 (2003).

[4]. V.G. Harris et al., J. Magn. Magn. Mater. **321**, 2035 (2009).

[5] P.P. Hankare, R. P. Patil, U. B. Sankpal, S. D. Jadhav, I. S. Mulla, K. M. Jadhav, and B. K. Chougule, J. Mag. Mag. Mater. **321**, 3270 (2009).

[6] M. P. Horvath, J. Magn. Magn. Mater. **215**, 171 (2000).

[7] Y. P. Fu, and C. S. Hsu, Solid State Comm.**134**, 201 (2005).

[8] G. Vijayasri, and R. N. Bhowmik, AIP Conf. Proc. **1512**, 1196 (2013).

[9] M. Deepa, P.P. Rao, A.N. Radhakrishnan, K.S. Sibi, and P. Koshy, Mater. Res. Bull. **44**, 1481 (2009).

[10] Li Lv, J.-P. Zhou, Q. Liu, G. Zhu, X. -Z. Chen, X. –B. Bian, and P. Liu, Physica E **43**, 1798 (2011).

[11] C. Brosseau, and W. N. Dong, J. Appl. Phys.**104**, 064108 (2008).

[12] C. Wende, and H. Langbein, Cryst. Res. Technol. **41**, 18 (2006).

[13] M. Younas, M. Nadeem, M. Atif, and R. Grossinger, J. Appl. Phys.**109**, 093704 (2011).

[14] A. U. Rahman, M. A. Rafiq, S. Karim, K. Maaz, M. Siddique, and M. M. Hasan, J. Phys. D: Appl. Phys.**44**, 165404 (2011).

[15] R. Kannan, S. Rajagopan, A. Arunkumar, D. Vanidha, and R. Murugaraj, J. Appl. Phys. **112**, 063926 (2012).

[16] I.P. Muthuselvam, and R.N. Bhowmik, J. Phys. D: Appl. Phys. **43**,465002 (2010).

[17] M. Hövel, B. Gompf, and M. Dressel, Thin Solid Films **519**, 2955(2011).

[18] N. V. Smith, Phys. Rev. B **64**, 155106 (2001).

[19] K. Lee, S. Cho, S. H. Park, A. J. Heeger, C. W. Lee and S. H. Lee, Nature **441**, 04705 (2006).

[20] J. B. Goodenough, Phys. Rev. **117**, 1442 (1960).

Figure captions:

Fig.1 (Colour online) Full Profile Fit of the XRD pattern of different annealed samples.

Fig. 2 SEM pictures for the samples S_8 (a), S_9 (b), S_10 (c), S_11 (d) and magnified form of surface strain (e). The EDX spectrum of sample S_11 is shown in (f).

Fig. 3 (Colour online) Frequency dependent ac conductivity ($\sigma_n(\nu)$) (normalized by the conductivity at 1 Hz ($\sigma_{1\,Hz}$) for samples S8 (a) and S9 (b). The figures are magnified in (c) and (d) to clearly show the semiconductor to metallic transition. The lines in (c-d) are the fitted data according to Jonscher power law and Drude model for semiconductor and metallic regimes, respectively.

Fig. 4 (Colour online) (a) temperature dependence of exponent obtained from fit of conductivity data in semiconductor regime using Jonscher power law (left hand side) and relaxation time obtained from fit of data in metallic regime using Drude model (right hand side). Temperature dependence of conductivity at different frequencies for the samples S8 (b) and S9 (c).

Fig. 5 (Colour online) Frequency dependence of real part ($\varepsilon'$) (a-b) and imaginary part ($\varepsilon''$) (c-d) of dielectric constant at different measurement temperatures.

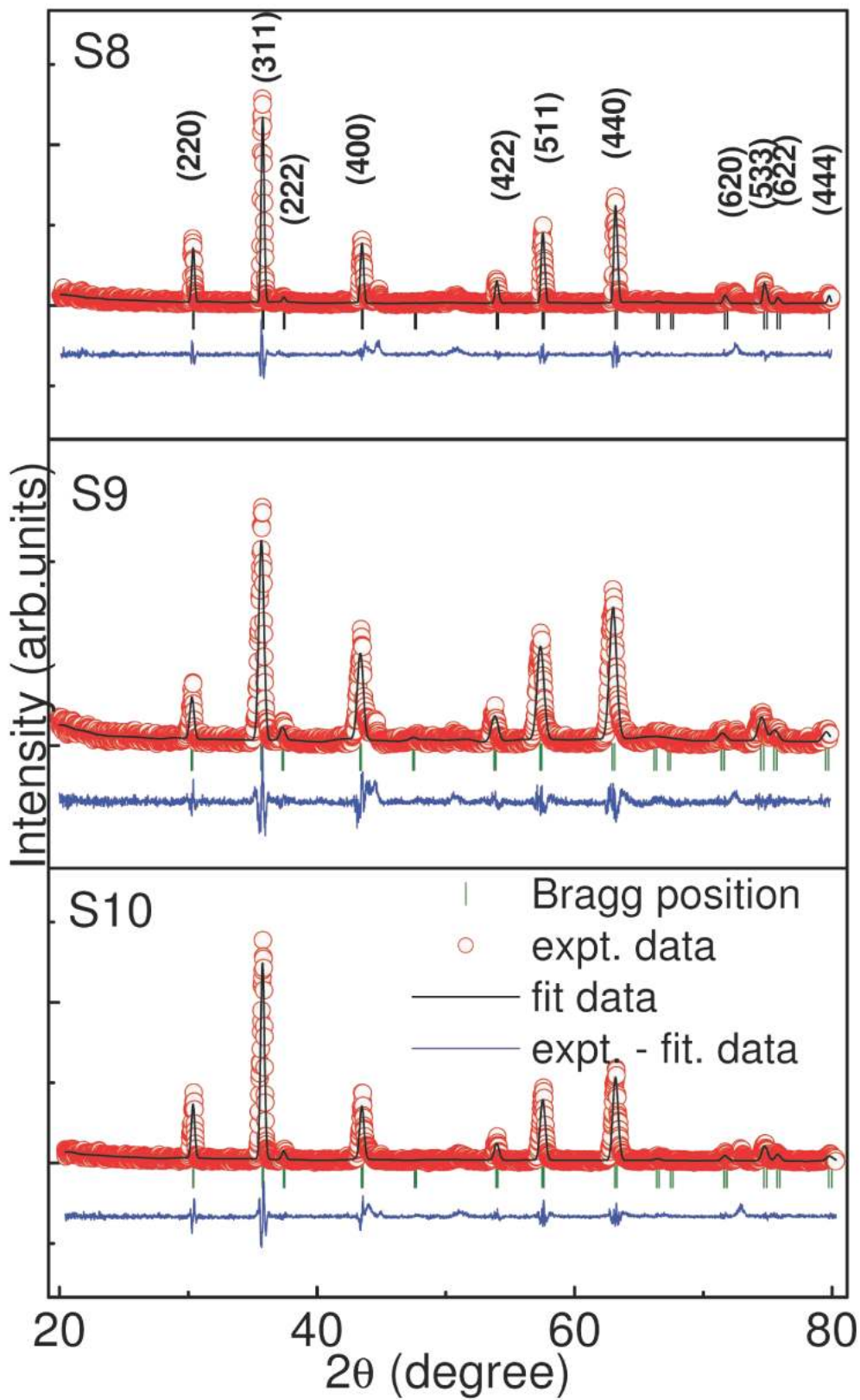

Fig.1 (Colour online) Full Profile Fit of the XRD pattern of different annealed samples.

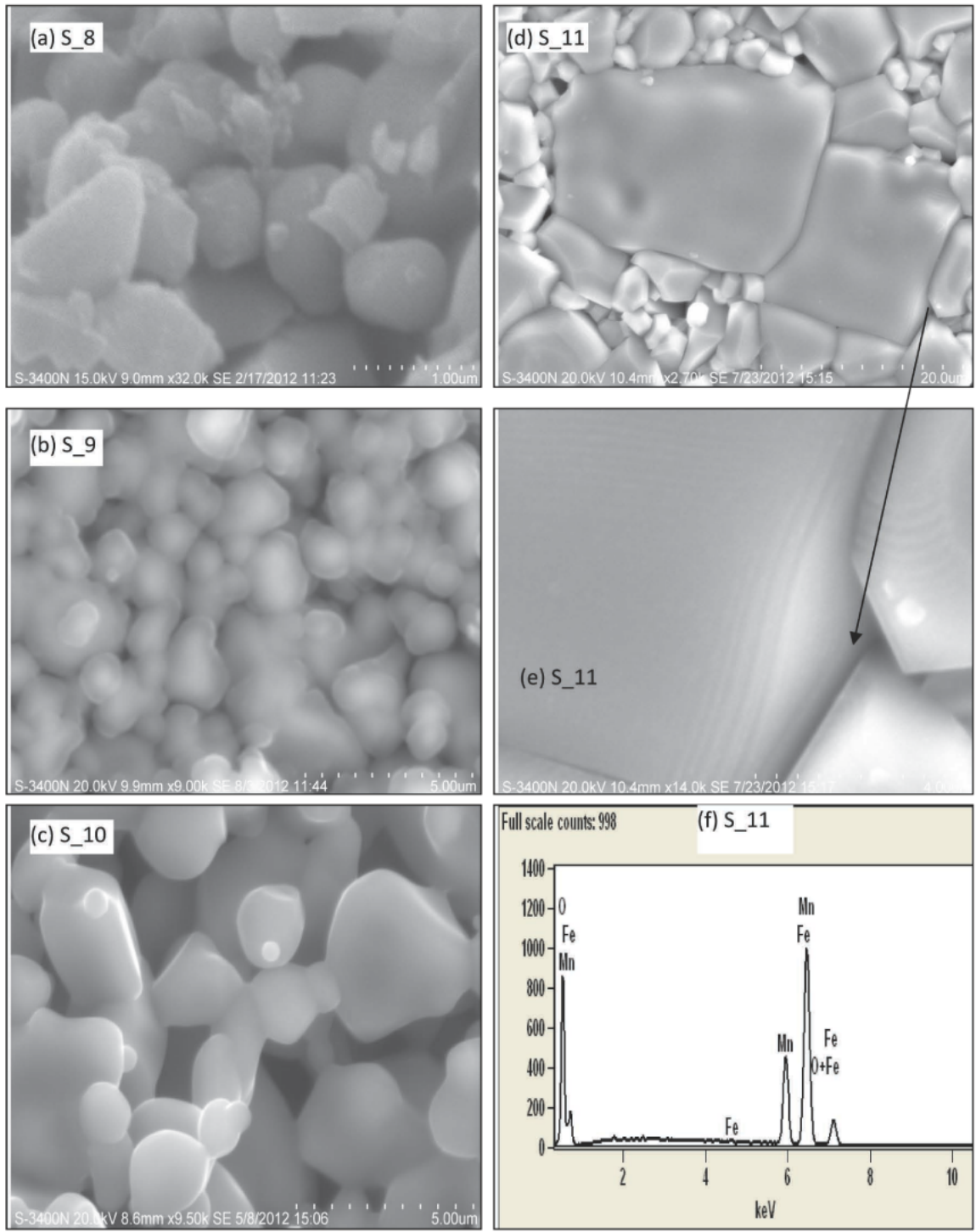

Fig. 2 SEM pictures for the samples S_8 (a), S_9 (b), S_10 (c), S_11 (d) and magnified form of surface strain (e). The EDX spectrum of sample S_11 is shown in (f).

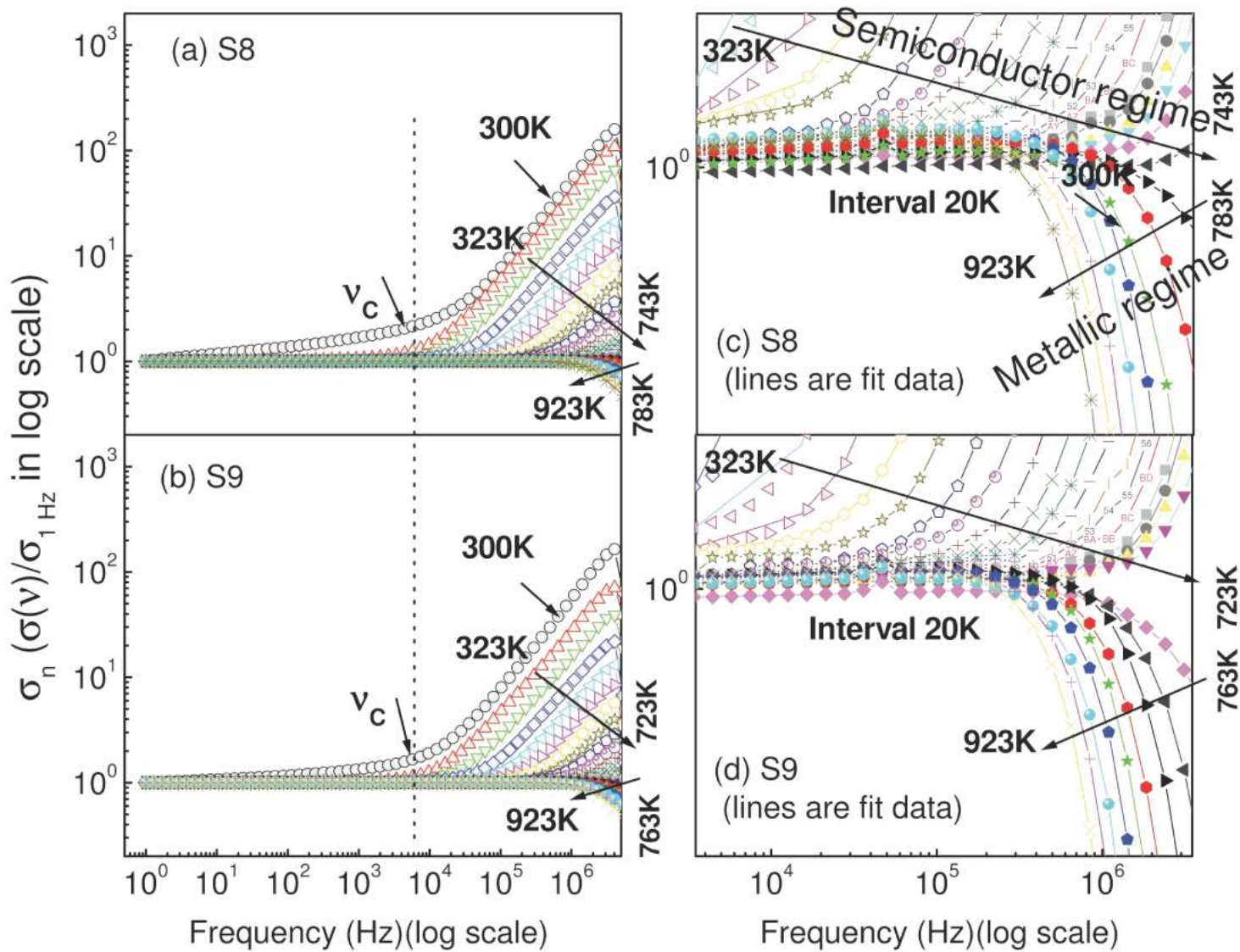

Fig. 3 (Colour online) Frequency dependent ac conductivity ($\sigma_n(\nu)$) (normalized by the conductivity at 1 Hz ($\sigma_{1\,Hz}$)) for samples S8 (a) and S9 (b). The figures are magnified in (c) and (d) to clearly show the semiconductor to metallic transition. The lines in (c-d) are the fitted data according to Jonscher power law and Drude model for semiconductor and metallic regimes, respectively.

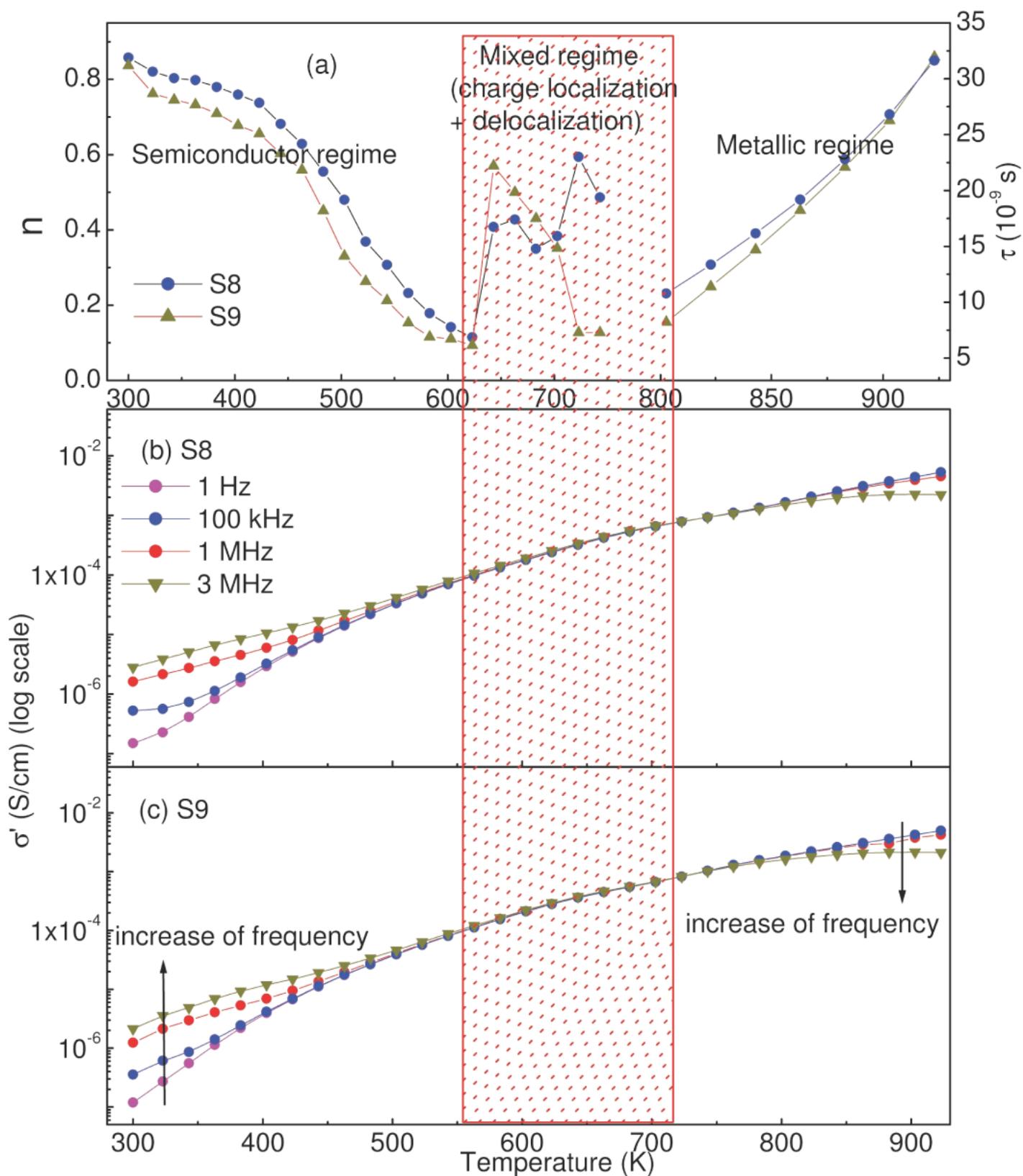

Fig. 4 (Colour online) (a) temperature dependence of exponent obtained from fit of conductivity data in semiconductor regime using Jonscher power law (left hand side) and relaxation time obtained from fit of data in metallic regime using Drude model (right hand side). Temperature dependence of conductivity at different frequencies for the samples S8 (b) and S9 (c).

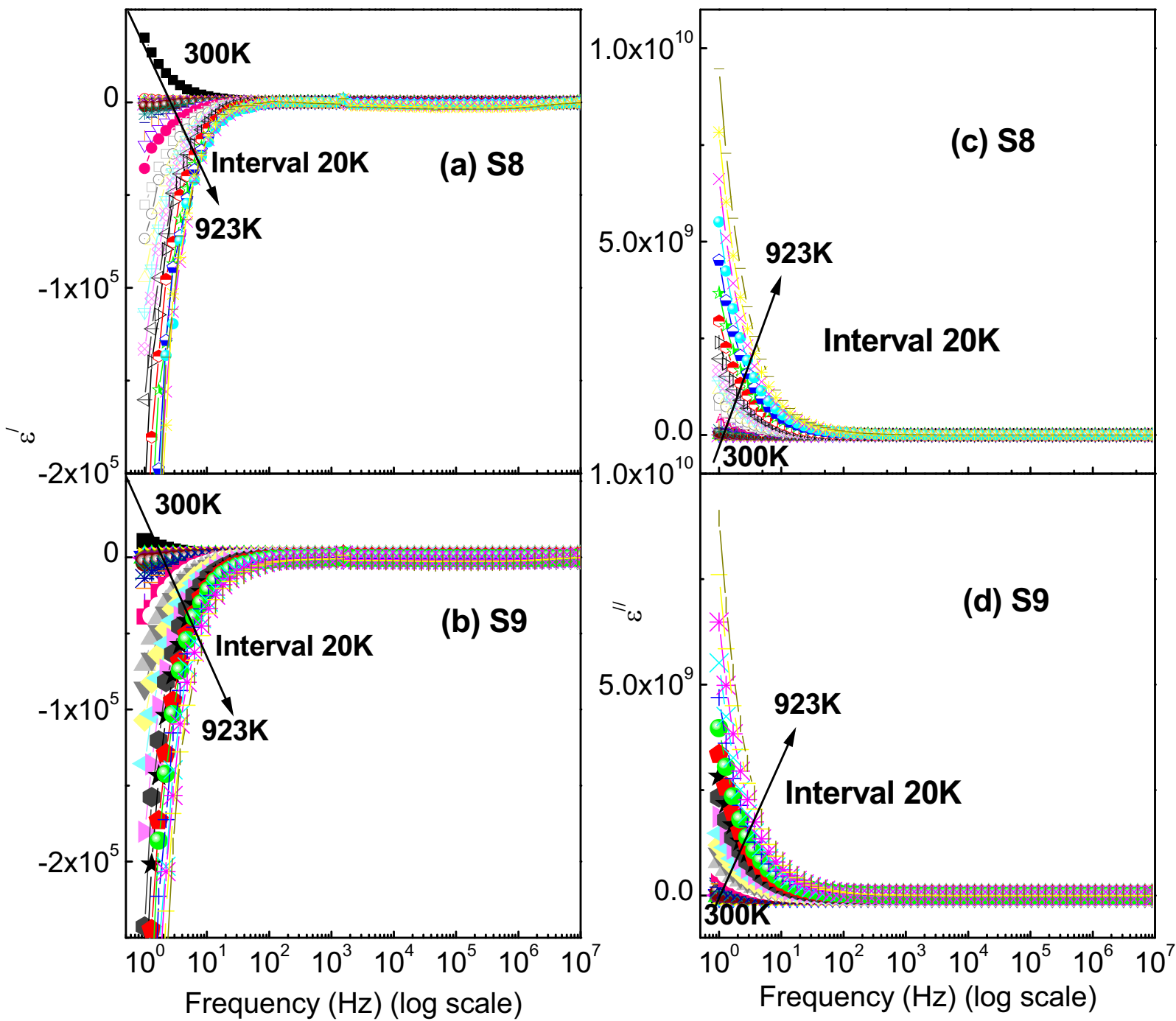

Fig. 5 (Colour online) Frequency dependence of real part ($\varepsilon'$) (a-b) and imaginary part ($\varepsilon''$) (c-d) of dielectric constant at different measurement temperatures.